# Some Analytical Results in (2+1) dimensional L.G.T.

M. Caselle[a]

[a]Dipartimento di Fisica Teorica dell'Università di Torino and INFN, Sezione di Torino,
V. P. Giuria 1, I-10125 Turin, Italy

We show that, within the framework of suitably chosen approximate effective actions it is possible to evaluate analytically the string tension, the spacelike string tension and the deconfinement temperature of (2+1) dimensional lattice gauge theories. In the case of SU(2) gauge theory our results agree with those obtained through Montecarlo simulations.

Lattice Gauge Theories (LGT) in (2+1) dimensions offer a unique tool to improve our understanding of the non perturbative behaviour of gauge theories. In fact they share with the (3+1) dimensional theories several non-perturbative features like confinement and the presence of a finite temperature deconfinement transition, but they are much simpler to study. In particular, in some suitable limits of the parameter space, one can construct approximate, effective actions which are exactly solvable. Their predictions can then be carefully tested by using high precision Montecarlo simulations which in (2+1) dimension are rather fast and do not require too large computers. This report is a short version of [1–3]. In particular we shall concentrate here on the SU(2) case, but interesting results can also be also obtained in the $Z_2$ (Ising gauge) [1] case and in the large $N$ limit [3].

## 1. General setting and numerical results

Let us consider a pure gauge theory with gauge group $SU(N)$, defined on a $2+1$ dimensional cubic lattice of $N_t$ ($N_s$) spacings in the time (space) direction. Gauge fields are described by the link variables $U_{n;i} \in SU(N)$, where $n \equiv (\vec{x}, t)$ denotes the space-time position of the link and $i$ its direction. We impose periodic boundary conditions in the time direction in order to describe a finite temperature LGT. Let us choose the standard Wilson action, but with different bare couplings in the time and space directions. Let us call them $\beta_t$ and $\beta_s$ respectively, and let us similarly call $S_s$ ($S_t$) the spacelike (timelike) part of the Wilson action. $\beta_s$ and $\beta_t$ are related to the (bare) gauge coupling $g$ and to the temperature $T$ by the relations

$$\frac{2N}{g^2} = a\sqrt{\beta_s \beta_t} \quad , \qquad T = \frac{1}{N_t a}\sqrt{\frac{\beta_t}{\beta_s}} \quad , \qquad (1)$$

where $a$ is the spacelike lattice spacing. The peculiar feature of the (2+1) dimensional case with respect to the (3+1) dimensional one is that the coupling constant $g^2$ has the dimension of a mass and sets the overall mass scale for all physical quantities. So near the continuum limit, dimensional quantities, like $\sqrt{\sigma}$ or $T_c$, can be written as power series of $g^2$. In the SU(2) case the first coefficients of these series are known with high precision (see [4] for details). In listing these results we shall fix $\beta_s = \beta_t = \beta = 4/(a\, g^2)$, in order to make contact with [4]. The zero temperature string tension $\sigma(0)$ and the deconfinement temperature $T_c$ behave as follows:

$$a\sqrt{\sigma(0)} = \frac{1.336(10)}{\beta} + \frac{1.122}{\beta^2} \quad , \qquad (2)$$

$$a\, T_c = \frac{1.50(2)}{\beta} \quad . \qquad (3)$$

In the deconfined phase $T > T_c$ the spacelike string tension rises linearly with the temperature according to the law:

$$a^2 \sigma_s(aT) = 1.46(8)\frac{aT}{\beta} \quad . \qquad (4)$$

These are the numbers which we shall try to obtain analytically in the following.

## 2. Analytic results.

All the approximations and the effective actions which we shall discuss in this section are rather well known, so we shall only state the results, and refer the reader to [1–3] where further details and reference to the original papers are given.

By means of an approximate renormalization group transformation, within the framework of a Migdal-Kadanoff bond-moving scheme, it is possible to reduce the original $(2 + 1)$ dimensional Wilson action, to a two-dimensional gauge theory coupled to the Polyakov lines

$$V(\vec{x}) = \prod_{t=1}^{N_t} U_{\vec{x},t;0} \quad , \tag{5}$$

which play the role of a Higgs field. While the spacelike part of the action remains unchanged, the timelike part, after a suitable truncation, becomes:

$$S_t = \sum_{\vec{x}} \frac{\hat{\beta}_t}{N} Re \sum_{i=1}^{2} Tr[V(\vec{x})U_{\vec{x};i}V^{\dagger}(\vec{x}+\hat{i})U_{\vec{x};i}^{\dagger}] \tag{6}$$

The net effect of the Migdal-Kadanoff approximation is to reduce the lattice size in the time direction $N_t$ to its extreme value $N_t = 1$, and to change (in the large $\beta_t$ limit in which we are interested) the couplings according to:

$$\hat{\beta}_t \sim \frac{\beta_t}{N_t} \quad , \qquad \hat{\beta}_s \sim \beta_s N_t \quad . \tag{7}$$

Let us look now to the high temperature, deconfined, phase of the model. In this phase the Polyakov loop has a non-zero expectation value, and it is an element of the center of the gauge group. It can be expanded around the vacuum as follows

$$V(\vec{x}) \equiv e^{i\frac{\phi(\vec{x})}{\sqrt{\beta_t}}} = 1 + i\frac{\phi(\vec{x})}{\sqrt{\beta_t}} - \frac{\phi^2(\vec{x})}{2\beta_t} + \cdots \tag{8}$$

where $\phi(x)$ is a Hermitian $N \times N$ matrix. By inserting (8) in $S_t$ we find:

$$\begin{aligned} S_t &= \frac{1}{N} Tr \sum_{\vec{x}} (-m_0^2 \phi(\vec{x})^2 \\ &+ \sum_{i=1}^{2} U_{\vec{x};i} \phi(\vec{x}) U_{\vec{x};i}^{\dagger} \phi(\vec{x}+\hat{i})) \quad , \end{aligned} \tag{9}$$

with $m_0^2 = d$. This approximation scheme for high temperature QCD is known in the literature as "complete dimensional reduction". Its major drawback is that (as a consequence of the Migdal Kadanoff transformation) it neglects the contributions from the non-static modes in the compactified time direction. However, in the high temperature limit, it can be shown (at one loop level) that such contributions only amount to a shift in the mass of the scalar field. All other contributions, like quartic self-interaction terms decrease with the temperature and can thus be neglected at sufficiently high temperature.

Since in $S_t$ the field $\phi$ appears only quadratically (even after the mass shift due to the non-static modes), it can be integrated exactly, leading to the so called "induced action" which is function of the spacelike gauge fields only [5],

$$S_{ind}[U] = -\frac{1}{2} \sum_{\Gamma} \frac{|TrU[\Gamma]|^2}{l[\Gamma](2m^2)^{l[\Gamma]}} \quad , \tag{10}$$

where $l[\Gamma]$ is the length of the loop $\Gamma$, $U[\Gamma]$ is the ordered product of link matrices along $\Gamma$ and the summation is over all closed loops. If we are interested to evaluate spacelike observables like the spacelike string tension, we can treat this induced action as a perturbation of the spacelike action $S_s$. Even if it looks rather complicated, $S_{ind}$ is all the same a two dimensional gauge theory, so we expect strong coupling expansion to be very effective in dealing with it. This is indeed the case and one can show explicitly that all contributions of order $1/\beta_s$ coming from $S_{ind}$ cancel exactly in the strong coupling expansion and that one is left with subleading $1/\beta_s^2$ contributions only [2].

As a consequence of this result, at the first order in $1/\beta_s$, the expectation value of the spacelike string tension is given by the spacelike part of the action only. This can be evaluated exactly and gives [2]):

$$a^2 \ \sigma(aT) = -\log\left(\frac{I_2(\hat{\beta}_s)}{I_1(\hat{\beta}_s)}\right) \tag{11}$$

where $I_n(\beta)$ is the $n^{th}$ modified Bessel function. By using the large $\beta$ expansion of the Bessel func-

tion and the relation between $\hat{\beta}_s$ and $\beta_s$, we obtain

$$a^2 \ \sigma(aT) = \frac{3 \ aT}{2\beta} + \cdots \quad , \qquad (12)$$

which is in remarkable agreement with the known value given in eq.(4). Eq.(12) can be easily generalized to any value of $N$:

$$a^2 \ \sigma(aT) = \frac{(N^2 - 1) \ aT}{2\beta} + \cdots \qquad (13)$$

Notice that within our framework the linear rise of the spacelike string tension with the temperature also has a natural explanation, since it simply encodes the $T$-dependence of $\beta_s$ with respect to $\beta$.

It is rather interesting to notice that, according to the Montecarlo simulations [4], this linear rise of the spacelike string tension sets in already at $T = T_c$, thus suggesting that the whole deconfined phase could be described using dimensional reduction. This leads to the following speculation. Let us assume an idealized picture in which the spacelike string tension is exactly constant as a function of $T$ in the region $T < T_c$, and then at $T = T_c$ sharply starts to rise linearly. This implies the relation (for a generic $N$):

$$\sigma(0) = aT_c\sigma(aT=1) = \frac{(N^2-1)T_c}{2a\beta} \qquad (14)$$

Let us further assume the following adimensional relation [6]

$$\sigma(0) = \frac{\pi}{3}T_c^2 \qquad (15)$$

which comes from the completely different context of the effective string approach to the interquark potential in LGT's. This relation does not depend on the gauge group, hence it holds unchanged for all $N's$. By combining (14) and (15) we can predict the value of the (zero temperature) string tension and the critical temperature for SU(N) gauge theories in (2+1) dimension as a function of $\beta$.

$$\sigma(0) = \frac{3 \ (N^2 - 1)^2}{4\pi\beta^2} \qquad (16)$$

$$T_c = \frac{3 \ (N^2 - 1)}{2\pi\beta} \qquad (17)$$

Needless to say, these results are not as reliable as those of eq.s (12) and (13), since both assumptions eq.s (14) and (15) only rely on qualitative arguments. Notwithstanding this, if we set $N = 2$ in the above eq.s (16) and (17) we find estimates for $\sqrt{\sigma}$ and $T_c$ which are only 10% and 5% far from the values reported in eq.s (2) and (3).

In view of this qualitative agreement, it would be important to have some independent check of these predictions. An interesting possibility is to look to the large $N$ limit, where mean field is exact, and reliable informations on the critical deconfinement temperature can be obtained. We studied recently this problem in [3], by combining techniques typical of the large $N$ matrix models and strong coupling expansions. In this limit eq.(17) gives:

$$\frac{T_c}{g^2} \equiv \frac{\beta_c}{N^2 \ N_t} = \frac{3}{2\pi} \sim 0.477 \quad . \qquad (18)$$

With our techniques we can give an upper bound for the large $N$ limit of the critical temperature, and suggest a tentative lower bound (which could still be affected by the contribution of the spacelike part of the action, see [3]). They are summarized as follows:

$$0.321 < \frac{T_c}{g^2} < 0.601 \qquad (19)$$

and are in agreement with eq.(18).